\documentstyle[12pt,epsf]{article}

\def\half{{\textstyle{1\over 2}}}

\def\sixth{{\textstyle {1\over6}}}

\def\inbar{\,\vrule height1.5ex width.4pt depth0pt}

\def\IC{\relax\hbox{$\inbar\kern-.3em{\rm C}$}}
\def\IQ{\relax\hbox{$\inbar\kern-.3em{\rm Q}$}}
\def\IR{\relax{\rm I\kern-.18em R}}
 \font\cmss=cmss10 \font\cmsss=cmss10 at 7pt

\def\IZ{\relax\ifmmode\mathchoice
 {\hbox{\cmss Z\kern-.4em Z}}{\hbox{\cmss Z\kern-.4em Z}}
 {\lower.9pt\hbox{\cmsss Z\kern-.4em Z}}
 {\lower1.2pt\hbox{\cmsss Z\kern-.4em Z}}\else{\cmss Z\kern-.4em Z}\fi}

\def\IZT{\IZ_2\times\IZ_2}

\def\Io{\relax\ifmmode\mathchoice
 {\hbox{\cmss 1\kern-.4em 1}}{\hbox{\cmss 1\kern-.4em 1}}
 {\lower.9pt\hbox{\cmsss 1\kern-.4em 1}}
 {\lower1.2pt\hbox{\cmsss 1\kern-.4em 1}}\else{\cmss 1\kern-.4em 1}\fi}

\def\cos{ {\rm cos}}
\def\sin{ {\rm sin}}
\def\exp{ {\rm exp}}

\begin{document}

\begin{titlepage}
\samepage{
\setcounter{page}{1}
\rightline{BU-HEPP-02/08}
\rightline{CTP-TAMU-02/03}
\rightline{\tt hep-ph/0210093n}
\rightline{October 2002}
\vfill
\begin{center}
 {\Large \bf Parameter Space Investigations of\\ 
             Free Fermionic Heterotic Models}
\vfill
\vskip .3truecm
\vfill {\large
        Gerald B. Cleaver,\footnote{Gerald{\underline{\phantom{a}}}Cleaver@baylor.edu}}
\vspace{.12in}

{\it        Center for Astrophysics, Space Physics \& Engineering Research\\
            Department of Physics, PO Box 97316, Baylor University\\
            Waco, Texas 97316, USA\\
            and\\
            Astro Particle Physics Group,
            Houston Advanced Research Center (HARC),\\
            The Mitchell Campus,
            Woodlands, TX 77381, USA\\}
\vspace{.025in}
\end{center}
\vfill
\begin{abstract}
I survey the parameter space of
NAHE-based free fermionic heterotic string models. First, I discuss flat
directions of the low energy effective field theories 
and show that $D$-flat directions need not be
isomorphic to gauge invariant superpotential terms.
Next, I review recent studies of three generation
$SU(3)_C\times SU(2)_L\times U(1)_Y$,
$SU(3)_C\times SU(2)_L\times SU(2)_R\times U(1)_{B-L}$,
$SU(4)_C\times SU(2)_L\times U(1)_{Y'}$, and
flipped $SU(5)$ models, constructed by Wilson breaking of the
underlying $SO(10)$ observable sector gauge group of the NAHE $\IZT$ basis.\\
{\it Based on talk presented at the First International Conference on String
Phenomenology.}
\end{abstract}
\smallskip}
\end{titlepage}

\setcounter{footnote}{0}

\section{NAHE-based $\IZ_2\times\IZ_2$ Heterotic Models}

The NAHE-based \cite{nahe} free fermionic \cite{ff} approach to 
heterotic superstring model building has proven very 
successful phenomenologically. This construction yielded 
Flipped SU(5) \cite{fsu5} 
around 1988, several standard-like models in the 1990's 
 \cite{afsl}, 
the Minimal Supersymmetric Heterotic Standard Model (MSHSM) \cite{cfnw} 
in 1999-2000, and in 2002 a standard-like model with the potential for 
optical unification \cite{reu}. 
%In this talk I review ongoing investigations of the 
%parameter space of this model class.

The NAHE set of basis vectors is composed of 5 sectors, 
$\{ {\bf 1}, {\bf S}, {\bf b}_1, {\bf b}_2, {\bf b}_3\}$. 
The antiperiodic sector ${\bf 0} \equiv {\bf 1}+{\bf 1}$ yields
the massless generators of the gauge group $SO(10)\times SO(6)^3 \times E_8$,
and ${\bf S}$ generates $N=1$ spacetime supersymmetry (SUSY), 
while each ${\bf b}_i$ produces 2 copies of 
$(\bf{16},\bf{4})$ and $(\bf{16},\overline{\bf{4}})$ of 
$SO(10)\times SO(6)_i$, respectively.  
Thus, the NAHE set contains $2\times (4+4)\times 3= 48$ generations of 
$SO(10)$. Additional sectors, acting like Wilson loops, 
can reduce both the gauge group and the 
number of copies of each generation. 
 For (quasi)-realistic models, 
$SO(10)$ is broken to a smaller gauge group containing
$SU(3)_C \times SU(2)_L \times U(1)_Y \times U(1)_{Z'}$,
where $U(1)_Y = U(1)_C + U(1)_L$ and $U(1)_{Z'} = U(1)_C - U(1)_L$, 
while
each ${\bf b}_{i}$ is constrained to produce only a single copy 
of a generation. This reduction occurs at the expense 
of many new Minimal Supersymmetric Standard Model (MSSM) charged 
exotics appearing from Wilson loop sectors. Further, an extra $U(1)$
usually becomes anomalous. This $U(1)_A$ has its origin either 
from $E_6$ decomposition into $SO(10)\times U(1)$ or from one of the 
$U(1)$'s generated by $D=10 \rightarrow D =4$ compactification. 
The two distinct $U(1)_A$ types have different 
phenomenological implications.

\section{Flat Directions in the Low Energy Effective Field Theory}

For quasi-realistic models, Wilson lines can break $SO(10)$ into
$SU(5)\times U(1)$,
$SO(6)\times SO(4)\sim SU(4)_C \times SU(2)_L \times SU(2)_R$, 
$SU(4)_C \times SU(2)_L \times U(1)_Y$,
$SU(3)_C \times SU(2)_L \times SU(2)_R \times U(1)_{B-L}$, or
$SU(3)_C \times SU(2)_L \times U(1)_{Y}$.
These gauge groups are found in models at various points
in the NAHE-based free fermionic parameter space. In the low energy 
effective field theory it is possible to move away from free fermionic
points through vacuum expectation values (VEVs) of scalar fields.
When an Abelian anomaly exists, 
the Green-Schwarz-Dine-Seiberg-Witten mechanism 
eliminates it and in the process generates a Fayet-Iliopoulos (F-I) $D$-term,
$\xi={{g^2({\rm Tr} Q_A)}\over{192\pi^2}}M_{\rm Pl}^2$.
$\xi$ imposes a VEV at a scale 
near $\sim \textstyle{1\over10} M_{\rm string}$ 
through the $D$-flatness constraints,
$D_A\equiv \sum_m Q_A^m\vert \langle\phi_m\rangle\vert^2+\xi = 0$ 
(for the anomalous Abelian symmetry), 
$D_i\equiv \sum_m Q_A^m\vert\langle\phi_m\rangle\vert^2 = 0$ 
(for the non-anomalous Abelian symmetries), and
$D_a^{\alpha}\equiv \sum_m \langle\phi_m^\dagger T^{\alpha}_a \phi_m\rangle = 0$ 
(for the non-Abelian symmetries).
For anomaly-free models, a $D$-flat VEV is perturbatively allowed, but its
scale is not specified.
A $D$-flat VEV must also be $F$-flat
%, i.e., $F_{\phi} \equiv \langle \partial W/\partial \phi \rangle = 0$ 
 to at least $17^{\rm th}$ order  
in $W$ for SUSY to remain unbroken down to $\sim 10$ TeV.   

For VEVs involving only non-Abelian singlets, all $D$-flat directions
correspond to gauge invariant superpotential terms. 
In ref.\ \cite{geom}
%``Geometrical Interpretations of
%non-Abelian $D$- and $F$-flat Direction Constraints'' \cite{geom} 
we show
that flatness is much more involved for non-Abelian fields, 
specifically that $D$-flat directions do not necessarily correspond to gauge 
invariant superpotential terms. Take, for example, $SU(2)$ doublets, 
$L_i$, where $R_i\cos \textstyle{\theta_i\over 2} 
\exp \{\textstyle{-i \phi\over 2}\}$ 
and $R_i\sin \textstyle{\theta_i\over 2} 
\exp \{\textstyle{+i \phi\over 2}\}$
are the ``up'' and ``down''  VEV components.
Then a D-flat solution can be formed from three doublets
$L_{i=1,2,3}$ of equal norms, 
$\vert R_1 \vert^2 = \vert R_2 \vert^2 = \vert R_3 \vert^2$, when
$\theta_1= 0, \theta_2= \theta_3 = \textstyle{2\pi\over3}$ and
$\phi_2= 0$, $\phi_3= \pi$. In contrast, the only superpotential
term containing three unique doublets is (up to an overall power) 
$L_1 L_2 (L_3)^2 \equiv (L_1\cdot L_3)(L_2\cdot L_3)$, which 
corresponds to the norms       
$\vert R_1 \vert^2 = \vert R_2 \vert^2 = 
\textstyle{1\over 2}\vert R_3 \vert^2$. 
For $D$-flat VEVs lacking superpotential correspondence, 
we present a geometric language for which there is intuitive and immediate
recognition of when $D$- and $F$-conditions are simultaneously compatible.
Our method provides a tool for comprehensive classification of flat 
directions. 

\section{Optical Unification}

An enduring issue in string theory has been the discrepancy between the
$SU(3)_C\times SU(2)_L\times U(1)_Y$ ([321])
gauge coupling unification scale, 
$\Lambda_U \approx 2.5 \times 10^{16}$ GeV \cite{mssmgcu}, 
for the the MSSM with intermediate scale desert
and the string scale, $\Lambda_H \approx 5\times 10^{17}$ GeV \cite{kapl}, 
for the weakly coupled heterotic string.

One explanation is that intermediate scale exotics could shift
the MSSM unification scale upward to the string scale \cite{mup}.
%The near ubiquitous appearance of MSSM-charged exotics in
%heterotic string models adds weight to this proposal. 
If MSSM exotics exist with
intermediate scale masses of order $\Lambda_I$, then the actual
[321] running couplings are altered above $\Lambda_I$. It is then,
perhaps, puzzling that the illusion of MSSM unification
should still be maintained when the
intermediate scale MSSM exotics are ignored \cite{jg1}.
Maintaining this illusion
likely requires very fine tuning of $\Lambda_I$ for a generic
exotic particle set and $\Lambda_H$.
Slight shifting of $\Lambda_I$ would, with high probability, 
destroy appearances.
Thus, in some sense, the apparent MSSM unification below the
string scale might be viewed as accidental \cite{ghil,jg1}.

Joel Giedt recently discussed a mechanism, 
entitled ``optical unification,'' whereby the appearance of
a $\Lambda_U$ is not accidental \cite{jg1}.
Optical unification results in $\Lambda_U$ not disappearing
under shifts of $\Lambda_I$.
Instead, $\Lambda_U$ likewise shifts in value.
This effect is parallel to a virtual image always appearing
between a diverging lens and a real object.
%independent of the position of the lens or real object.
Hence, Giedt's choice of appellation for this mechanism.

Successful optical unification requires three things \cite{jg1}.
First, the effective level of the
hypercharge generator must be the standard
$k_Y = \textstyle{5\over3}$.
This is a
strong constraint on string-derived $[321]$ models,
for the vast majority have non-standard
hypercharge levels.
Only select classes of models, such as the NAHE-based 
free fermionic class,
can yield $k_{Y} = \textstyle{5\over3}$.
Second, optical unification imposes the relationship
$\delta b_2 = \textstyle{7\over12} \delta b_3 + \textstyle{1\over4} 
\delta b_Y$,
between the exotic particle contributions $\delta b_3$, $\delta b_2$, 
and $\delta b_1$
to the [321] beta function coefficients.
Each $SU(3)_C$ exotic triplet or anti-triplet contributes
$\half$ to $\delta b_3$;
each $SU(2)_C$ doublet contributes
$\half$ to $\delta b_2$.
With the hypercharge of a MSSM quark doublet normalized to $\sixth$,
the contribution to $\delta b_Y$ from an individual particle with
hypercharge $Q_Y$ is $Q_Y^2$.
$\delta b_3 > \delta b_2$ is required
to keep the virtual unification scale below the string scale.
Combining this with the second constraint imposes
$\delta b_3 >  \delta b_2 \ge \textstyle{7\over12} \delta b_3$,
since $\delta b_Y \geq 0$.

To acquire intermediate scale mass,
the exotic triplets and anti-triplets must be equal in number.
Similarly, the exotic doublets must be even in number.
Hence, $\delta b_3$ and $\delta b_2$ must be integer \cite{jg1}.
As Giedt pointed out, the simplest solution to optical unification
is a set of
three exotic triplet/anti-triplet pairs and two pairs of doublets.
One pair of doublets can carry $Q_Y=\pm \half$, while the remaining
exotics carry no hypercharge.
Alternately, if the doublets carry too little hypercharge,
some exotic $SU(3)_C \times SU(2)_L$ singlets could make up the
hypercharge deficit.
The next simplest
solution requires four triplet/anti-triplet pairs and three pairs of
doublets that yield $\delta b_Y = 2 \textstyle{2\over 3}$
either as a set, or with the assistance of additional non-Abelian singlets.
In ref.\ \cite{reu}, we present a standard-like model that has the potential 
to realize the latter optical unification solution, 
with the required hypercharge carried by the (anti)-triplets and 
one additional pair of singlets. 
Detailed analysis of this model is underway.

\section{Left-Right Symmetric Models}

Our parameter space investigations have led to the discovery of four models
with left-right symmetry, two lacking \cite{lrs} and two containing an 
anomalous $U(1)$ \cite{lrs,fdlrs}.
The first non-anomalous model has gauge group
$SU(3)_C \times SU(2)_L \times SU(2)_R \times U(1)_{B-L}
\times \prod_{i=1}^{6} U(1)_i \times SU(3)_{H_1} \times SU(3)_{H_2}
\times \prod_{i=7}^{10} U(1)_i$, while the second has both observable and hidden
sector symmetries enhanced by Wilson loops, 
 $SU(3)_C \times SU(2)_L \times SU(2)_R \times SU(2)_{B-L} \times U(1)_{Y'}
\times \prod_{i=1}^{7} U(1)_i \times SU(4)_{H_1} \times SU(4)_{H_2}
\times \prod_{i=8}^{9} U(1)_i$. The lack of an anomaly in these models is related to
neither model having its basis in $SO(12)\times E_8 \times E_8$ nor in 
$SO(12)\times SO(16) \times SO(16)$, but rather in 
$SO(16)\times E_7 \times E_7$.  

Being non-anomalous, these models have no inherent
F-I VEV scale. Both models contain a neutral SM singlet in a $\mathbf{16}$ 
of $SO(10)$ 
which can break $SU(2)_R$, but the corresponding component in a 
$\overline{\mathbf{16}}$ is 
absent. Instead the $\overline{\mathbf{16}}$ 
carries a $SU(3)_C \times SU(2)_L$-singlet with
non-zero electromagnetic charge. Further, all other $SU(2)_R$ 
exotic doublets carry fractional electromagnetic charge. 
Thus, $SU(2)_R$ cannot be broken along a SUSY flat direction in these two 
models. However, Model 2 contains $SU(2)_{B-L}$ doublets that break 
$SU(2)_{B-L}$ along flat directions.   

An interesting feature of Model 1 is that its 
entire hidden sector matter spectrum can 
get superheavy along some flat directions, while the hidden sector gauge 
bosons remain massless. Thus, the 
lightest hidden sector states can be glueballs that only interact with 
SM states via superheavy fermions. 
This makes for interesting dark matter candidates.
In both models, all mass couplings involve Higgs bi-doublets. This implies
a danger of inducing FCNC's at an unacceptable rate. 

The two anomalous models have gauge groups identical to Models 1 and 2,
respectively, except that $U(1)_1$ becomes $U(1)_A$. 
Here the anomaly does not arise from a $U(1)$ within $E_6$, 
but from six-dimensional compactification. 
Model 3 contains Higgs bi-doublets, while Model 4 does not. Model 4 also lacks
Abelian flat directions. In fact, the anomaly can only be cancelled by 
$SU(2)_L$ and/or $SU(2)_R$ doublets taking on VEVs. Thus, at least one of 
$SU(2)_L$ or $SU(2)_R$ must be broken in Model 4.

$SU(4)_C \times SU(2)_L \times U(1)_{Y'}$ models are also under investigation.
%A non-anomalous model with 
% $SU(4)_C \times SU(2)_L \times U(1)_{Y'}$$\times \prod_{i=1}^{6} U(1)_i$
%$\times\prod_{i=1}^{5} SU(2)_{H_i}$$\times SU(4)_H$
%has been constructed, as has an anomalous model with 
% $SU(4)_C \times SU(2)_L \times U(1)_{Y'}
%\times U(1)_A \times \prod_{i=1}^{7} U(1)_i 
%\times SU(2)_{H} \times SU(4)_H \times SO(5)_H$.
Detailed properties of these models will appear in an upcoming paper 
\cite{421m}.      

\section{Flipped $SU(5)$}

Heterotic flipped $SU(5)$ had its origin back in
the late 1990's \cite{fsu5} and is the most developed of all string models. 
The complete gauge group for this model is
$\tilde{SU}(5) \times U(1) $$\times \prod_{i=1}^{4} U(1)_i 
$$\times SO(10)_H \times SO(6)_H$.
The hidden sector states are the
$T_{i= 1\  {\rm to}\ 5}$ ($\bf{10}$ reps of $SO(10)_H$), 
$\Delta_{i= 1\  {\rm to}\ 5}$ ($\bf{6}$ reps of $SO(6)_H$),
and the  
$\tilde{F}_{j= 1\  {\rm to}\ 6}$ and 
$\tilde{\overline{F}}_{j= 1\  {\rm to}\ 6}$,   
($\bf 4$ and $\overline{\bf 4}$ reps of $SO(6)_H$, respectively). 
Flat directions for $U(1)_A$ cancellation generally induce 
near string-scale mass for some, 
but not all, of these states \cite{gen1}. Depending on the number of 
$T_i$ and $\Delta_i$ remaining massless,  
the $SO(10)$ condensate scale is around $10^{14\, {\rm to}\, 15}$ GeV 
and that for $SO(6)$ is around $10^{12\, {\rm to}\, 13}$ GeV.    
  
The $\tilde{F}_{3,5}$ and $\tilde{\overline{F}}_{3,5}$ states 
always remain massless down to the condensate scales, 
where they may form four-field ``tetron'' condensates \cite{nbe}.
Uncharged tetrons could make good dark matter candidates \cite{nbe}. 
Coriano et al.\  has recently raised the issue of 
the relative lifetimes of the electromagnetically charged to uncharged 
tetrons \cite{cfp}. 
The related decay terms for charged and uncharged tetrons are currently 
under investigation  \cite{fsu502}.  

\section{Conclusion}

The parameter space of NAHE-class  \cite{nahe} free fermionic \cite{ff} 
heterotic strings has, over the last 15 years, provided (quasi)-realistic 
models with very interesting combinations of phenomenological features. 
Many features of a given model are generally very dependent upon 
choice of perturbative flat direction. 
Some of the most desired properties, such as a standard hypercharge 
embedding, are very difficult to find in the parameter space of other model 
classes. The flipped $SU(5)$ model \cite{fsu5}, the MSHSM \cite{cfnw}, 
and the recently-discovered optical unification model \cite{reu}, 
have especially interesting properties. 
There is ongoing work to embed these models in M-theory \cite{embed}. 

\section*{Acknowledgments}
G.C. sincerely thanks his collaborators of the various papers discussed herein:
D. Clements, J. Ellis, A.E. Faraggi, E. Mayes,
D.V. Nanopoulos, S. Nooij, J. Perkins, C. Savage,
T. Velhuis, and J. Walker 
(listed alphabetically). 
The research leading to these papers was 
partially supported by DOE grant DE-FG-0395ER40917 (GC,DVN,JW).

%========================================================================
%          MACROS FOR REFERENCES
%========================================================================
\def\AEF{A.E. Faraggi}
\def\AP#1#2#3{{\it Ann.\ Phys.}\/ {\bf#1} (#2) #3}
\def\NPB#1#2#3{{\it Nucl.\ Phys.}\/ {\bf B#1} (#2) #3}
\def\NPBPS#1#2#3{{\it Nucl.\ Phys.}\/ {{\bf B} (Proc. Suppl.) {\bf #1}} (#2)
 #3}
\def\PLB#1#2#3{{\it Phys.\ Lett.}\/ {\bf B#1} (#2) #3}
\def\PRD#1#2#3{{\it Phys.\ Rev.}\/ {\bf D#1} (#2) #3}
\def\PRL#1#2#3{{\it Phys.\ Rev.\ Lett.}\/ {\bf #1} (#2) #3}
\def\PRT#1#2#3{{\it Phys.\ Rep.}\/ {\bf#1} (#2) #3}
\def\PTP#1#2#3{{\it Prog.\ Theo.\ Phys.}\/ {\bf#1} (#2) #3}
\def\MODA#1#2#3{{\it Mod.\ Phys.\ Lett.}\/ {\bf A#1} (#2) #3}
\def\MPLA#1#2#3{{\it Mod.\ Phys.\ Lett.}\/ {\bf A#1} (#2) #3}
\def\IJMP#1#2#3{{\it Int.\ J.\ Mod.\ Phys.}\/ {\bf A#1} (#2) #3}
\def\IJMPA#1#2#3{{\it Int.\ J.\ Mod.\ Phys.}\/ {\bf A#1} (#2) #3}
\def\nuvc#1#2#3{{\it Nuovo Cimento}\/ {\bf #1A} (#2) #3}
\def\RPP#1#2#3{{\it Rept.\ Prog.\ Phys.}\/ {\bf #1} (#2) #3}
\def\etal{{\it et al\/}}
%=========================================================================

\def\and{\& }

\end{document}